\documentclass[3p,sort&compress]{elsarticle}
\usepackage{amsmath}
\usepackage{amssymb}
\usepackage{bm}
\usepackage[newfloat=true,frozencache]{minted}
\usepackage[hypertexnames=false]{hyperref}

\usepackage[nameinlink]{cleveref}

\Crefname{figure}{Fig.}{Figs.}
\usepackage[mathscr]{eucal}

\usepackage[hyperref,usenames,dvipsnames]{xcolor}
\usepackage{xpatch}
\xapptocmd{\inputminted}{\\[-\baselineskip]\noindent}{}{} \usepackage{multirow}

\newdefinition{rmk}{Remark}

\definecolor{CeruleanRef}{RGB}{12,127,172}
\hypersetup{
colorlinks = true
}

\let\max\relax \DeclareMathOperator*\max{\vphantom{p}max}
\let\subset\relax \DeclareMathOperator{\subset}{\subseteq}
 
\let\tilde\widetilde
\let\hat\widehat

\newcommand\fnurl[2]{\href{#1}{#2}\footnote{\scalebox{.8}{\url{#1}}}}

\newmintedfile[matlabfile]{matlab}{
fontsize=\footnotesize,
numberblanklines=true,
numbersep=5pt,
gobble=0,
framerule=0.4pt,
framesep=2mm,
breaklines=true,
linenos,
bgcolor=black!4!white
}

\newcommand\Tstrut{\rule{0pt}{2.6ex}}         \newcommand\Bstrut{\rule[-0.9ex]{0pt}{0pt}}   

\newcommand{\sspace}{\hspace{0.25pt}}

\newcommand{\R}{\mathbb{R}}        \newcommand{\dd}{\,\mathrm{d}}

\newcommand{\bdelta}{{\bm{\delta}}}

\newcommand{\bvarphi}{\bm{\varphi}}

\newcommand{\bmx}{\bm{x}}

\renewcommand{\sfn}{\mathsf{n}}

\newcommand{\sfA}{\mathsf{A}}

\newcommand{\scD}{\mathscr{D}}

\begin{document}
\hypersetup{
allcolors=CeruleanRef
}
\begin{frontmatter}
\title{The surrogate matrix methodology: A reference implementation for low-cost assembly in isogeometric analysis}

\author{Daniel Drzisga\corref{cor1}}
\ead{drzisga@ma.tum.de}
\author{Brendan Keith}
\ead{keith@ma.tum.de}
\author{Barbara Wohlmuth}
\ead{wohlmuth@ma.tum.de}

\cortext[cor1]{Corresponding author}
\address{Lehrstuhl f\"ur Numerische Mathematik, Fakult\"at f\"ur Mathematik (M2), Technische Universit\"at M\"unchen, Garching bei M\"unchen}
\begin{abstract}
A reference implementation of a new method in isogeometric analysis (IGA) is presented.
It delivers low-cost variable-scale approximations (surrogates) of the matrices which IGA conventionally requires to be computed by element-scale quadrature.
To generate surrogate matrices, quadrature must only be performed on a fraction of the elements in the computational domain.
In this way, quadrature determines only a subset of the entries in the final matrix.
The remaining matrix entries are computed by a simple B-spline interpolation procedure.
We present the modifications and extensions required for a reference implementation in the open-source IGA software library GeoPDEs.
The exposition is fashioned to help facilitate similar modifications in other contemporary software libraries.
\\

\noindent\textit{Method name:} Surrogate matrix method for isogeometric analysis
\\[-1.5\baselineskip]
\end{abstract}
\end{frontmatter}

\medskip
\noindent
\begin{tabular*}{\textwidth}{l@{\extracolsep{\fill}\quad}ll}
  \textbf{Specification Table} \\
  \hline			
  Subject Area & Mathematics \Tstrut\\
  More specific subject area: & Computational science \& engineering\\
  Method name: & Surrogate matrix method for isogeometric analysis \\
  Name and reference of & \multirow{9}{12cm}{T. J. Hughes, J. A. Cottrell, Y. Bazilevs, Isogeometric analysis: CAD, finite elements, NURBS, exact geometry and mesh refinement, Comput. Methods Appl. Mech. Eng. 194 (2005) 4135–4195.\\
  S. Bauer, M. Mohr, U. R\"ude, J. Weism\"uller, M. Wittmann, B. Wohlmuth, A two-scale approach for efficient on-the-fly operator assembly in massively parallel high performance multigrid codes, Appl. Numer. Math. 122 (2017) 14–38.\\
  D. Drzisga, B. Keith, B. Wohlmuth, The surrogate matrix methodology: Low-cost assembly for isogeometric analysis, arXiv preprint (2019).}\\
  ~	original method: & \\
  \\
  \\
  \\
  \\
  \\
  \\
  \\
  Resource availability: & Source code (Matlab) for full method implementation in GitHub repository\\
  & (\href{https://dx.doi.org/10.5281/zenodo.3402341}{https://dx.doi.org/10.5281/zenodo.3402341}). \Bstrut\\
  \hline  
\end{tabular*}

\section*{Method details} In \cite{drzisga2019igasurrogate}, we applied the surrogate matrix methodology to isogeometric analysis (IGA) in order to avoid \emph{over-assembling} mass, stiffness, and divergence matrices.
While doing so, we also showed how this methodology could be applied to other system matrices arising in IGA computations.
In that article, large performance gains were demonstrated for a number of important PDE scenarios while still maintaining solution accuracy.
In the present article, we provide an accompanying reference implementation, together with explanations and examples.
This should be considered an extension to the main theoretical article \cite{drzisga2019igasurrogate}, which we strongly recommend to read beforehand.

Recall that \cite{drzisga2019igasurrogate} focuses on the Galerkin form of IGA \cite{cottrell2009isogeometric,hughes2005isogeometric}.
The resulting surrogate matrix methods perform quadrature for only a small fraction of the NURBS basis function interactions during assembly and then \emph{approximate the rest by interpolation}.
This leads to large sparse linear system matrices where the majority of entries have been computed by interpolation.
Such interpolated matrices will generally not coincide with those which would otherwise be generated by performing quadrature for the complete basis, or on every element, but they can be interpreted as cost-effective surrogates for them.

This idea was first introduced in the context of first-order finite elements and massively parallel simulations by Bauer et al. in \cite{bauer2017two}.
There, the computational run time improvement results from a two-scale strategy after which the method was named.
Thereafter, several subsequent investigations were initiated \cite{bauer2018large,bauer2018new,drzisga2018surrogate,drzisga2019igasurrogate}, of which this paper may be seen as a product.
In particular, massively parallel applications in geodynamical simulations are presented in \cite{bauer2018large,bauer2018new} and a theoretical analysis for first order finite element methods is given in \cite{drzisga2018surrogate}.
For moderately sized problems a two-scale strategy will possibly not result in an optimal performance gain.
Therefore, we exploit a mesh-size dependent macro-element approach where the macro-mesh is closer related to the fine mesh.

In this article, we restrict our attention to Poisson's boundary value problem on a single patch geometry $\Omega = \bvarphi(\hat{\Omega})$, where $\hat{\Omega} = (0,1)^n$, $\bvarphi:\hat{\Omega}\to \R^n$ is a fixed diffeomorphism of sufficient regularity, and $n=2,3$.
Represented on the reference domain $\hat{\Omega}$, the bilinear form appearing in the standard weak form of this problem is defined
\begin{equation*}
\hat{a}(\hat{w},\hat{v})
=
\int_{\hat{\Omega}} \hat{\nabla} \hat{w}(\hat{\bmx})^\top {K}(\hat{\bmx})\, \hat{\nabla} \hat{v}(\hat{\bmx}) \dd \hat{\bmx}
\,,
\qquad
{K} = \frac{D\bvarphi^{-1}\sspace\sspace D\bvarphi^{-\top}}{|\det{\left(D\bvarphi^{-1}\right)}|}
\,,
\end{equation*}
for arbitrary $\hat{w},\hat{v} \in H^1(\hat{\Omega})$.
Adopting further notation from \cite{drzisga2019igasurrogate}, this reference domain description leads to the definition of a set of smooth \emph{stencil functions} $\Phi_\bdelta\colon \hat{\Omega}\supsetneq\tilde{\Omega}\to \R$, $\bdelta\in\scD$.
In turn, these functions can be interpolated onto a finite-dimensional space of multivariate B-splines using their values sampled from a subset $\{\tilde{\bmx}_i^\mathrm{s}\}\subset \tilde{\Omega}\cap\{\tilde{\bmx}_i\}$ of a lattice $\{\tilde{\bmx}_i\}$.
Such interpolants are called \emph{surrogate stencil functions} $\tilde{\Phi}_\bdelta\colon \tilde{\Omega}\to \R$.
Taking on the convention $\bdelta = \tilde{\bmx}_j - \tilde{\bmx}_i$, we define the surrogate stiffness matrix $\tilde{\sfA} \in \R^{N\times N}$ as follows:
\begin{equation*}
\tilde{\sfA}_{ij}
=
\begin{cases}
\tilde{\Phi}_\bdelta(\tilde{\bmx}_i) & \text{if } \tilde{\bmx}_i,\tilde{\bmx}_j\in\tilde{\Omega} \text{ and } i<j,\\
\tilde{\sfA}_{ji} & \text{if } \tilde{\bmx}_i,\tilde{\bmx}_j\in\tilde{\Omega} \text{ and } i>j,\\
\sfA_{ij}  &\text{in all other cases where } i\neq j,\\
- \sum_{k\neq i}\tilde{\sfA}_{ik} & \text{if } i=j.\\
\end{cases}
\end{equation*}
This definition preserves the symmetry and the kernel of the true IGA stiffness matrix $\sfA \in \R^{N\times N}$.

The reference implementation described in this paper is built on the GeoPDEs package for Isogeometric Analysis in Matlab and Octave \cite{de2011geopdes,vazquez2016new}.
This package provides a framework for implementing and testing new isogeometric methods for PDEs.
Our reference implementation is available in the git repository \cite{githubdrzisga}, which is itself a fork of the GeoPDEs repository.
It is important to note that similar modifications can be made to other present-day software libraries and so the implementation presented in this article should foremost serve as a template.
For this reason, we tried to find a balance between comprehensibility and performance.
Most notably, the performance of our implementation may be greatly improved by performing quadrature for individual basis function interactions instead of element-wise quadrature.
However, for most IGA software libraries, this feature would require significant software restructuring.

Before continuing, let us establish some more mathematical notation which appears in the article, cf.~\cite{drzisga2019igasurrogate}.
Let $n$ be the space dimension, $p$ be the maximum degree of the discrete IGA B-spline spaces, and $h$ be the mesh size.
By $M$, we denote the skip parameter which defines the sampling length $H = M \cdot h$ on which we perform interpolation of the stencil functions.
Let $q$ be the spline degree of the interpolation space.
We assume that the number of elements in each spatial dimension is always equal and we denote its value by $N_\mathrm{el}$.
Finally, recall that B-spline and NURBS bases are built from tensor products of univariate B-splines.
For convenience, we assume that the associated number of knots and the B-spline degrees do not depend on the Cartesian coordinates.

\section*{Implementation}

Our implementation preserves the local element quadrature approach present in most standard IGA and finite element software.
This is not optimally efficient, but performance advantages can still be easily achieved because quadrature is usually not required on every element.
In order to avoid performing quadrature on specific elements, we made some minor modifications to the following core functions in GeoPDEs: \fnurl{https://github.com/drzisga/geopdes/blob/v3.1-surrogate/geopdes/inst/msh/\%40msh\_cartesian/msh\_evaluate\_col.m}{\texttt{msh\_evaluate\_col.m}}, \fnurl{https://github.com/drzisga/geopdes/blob/v3.1-surrogate/geopdes/inst/space/\%40sp\_scalar/sp\_evaluate\_col.m}{\texttt{@sp\_scalar/sp\_evaluate\_col.m}}, and \fnurl{https://github.com/drzisga/geopdes/blob/v3.1-surrogate/geopdes/inst/space/\%40sp\_scalar/sp\_evaluate\_col\_param.m}{\texttt{@sp\_scalar/sp\_evaluate\_col\_param.m}}.\footnote{The interested reader may refer to the diff between the surrogate and the GeoPDEs master branch in order to view the precise modifications.
}
In addition to these minor modifications, we added two functions which handle the assembly of the surrogate stiffness matrices in 2D and 3D, respectively.
Namely, we added \fnurl{https://github.com/drzisga/geopdes/blob/v3.1-surrogate/geopdes/inst/space/\%40sp\_scalar/op\_gradu\_gradv\_surrogate\_2d.m}{\texttt{op\_gradu\_gradv\_surrogate\_2d}} and \fnurl{https://github.com/drzisga/geopdes/blob/v3.1-surrogate/geopdes/inst/space/\%40sp\_scalar/op\_gradu\_gradv\_surrogate\_3d.m}{\texttt{op\_gradu\_gradv\_surrogate\_3d}}.
These functions are based on \fnurl{https://github.com/drzisga/geopdes/blob/v3.1-surrogate/geopdes/inst/space/\%40sp\_scalar/op\_gradu\_gradv\_tp.m}{\texttt{@sp\_scalar/op\_gradu\_gradv\_tp}} wherein the global matrix is assembled in a column-wise fashion \cite{vazquez2016new}.
In this section, we outline the modifications made to these core routines and describe the new 2D function in detail by breaking it down into coherent pieces of code.

\subsection*{Code modifications}
\label{sec:modifications}

In GeoPDEs, column-wise assembly which exploits the tensor product structure in IGA is used for performance \cite{vazquez2016new}.
For instance, in 2D, the functions listed above are called for each column of elements\footnote{
In 3D, each column in the first dimension corresponds to a plane in the remaining dimensions, thus multiple element masks are required to cover all the active elements.
} in a patch during the column-wise assembly of the global stiffness matrix.
Here, the function \texttt{msh\_evaluate\_col} collects the quadrature rules for the physical domain in a column and the function \texttt{sp\_evaluate\_col} returns a \texttt{struct} variable representing the discrete function space in that column.

\begin{figure}\centering
\includegraphics[height=5.5cm]{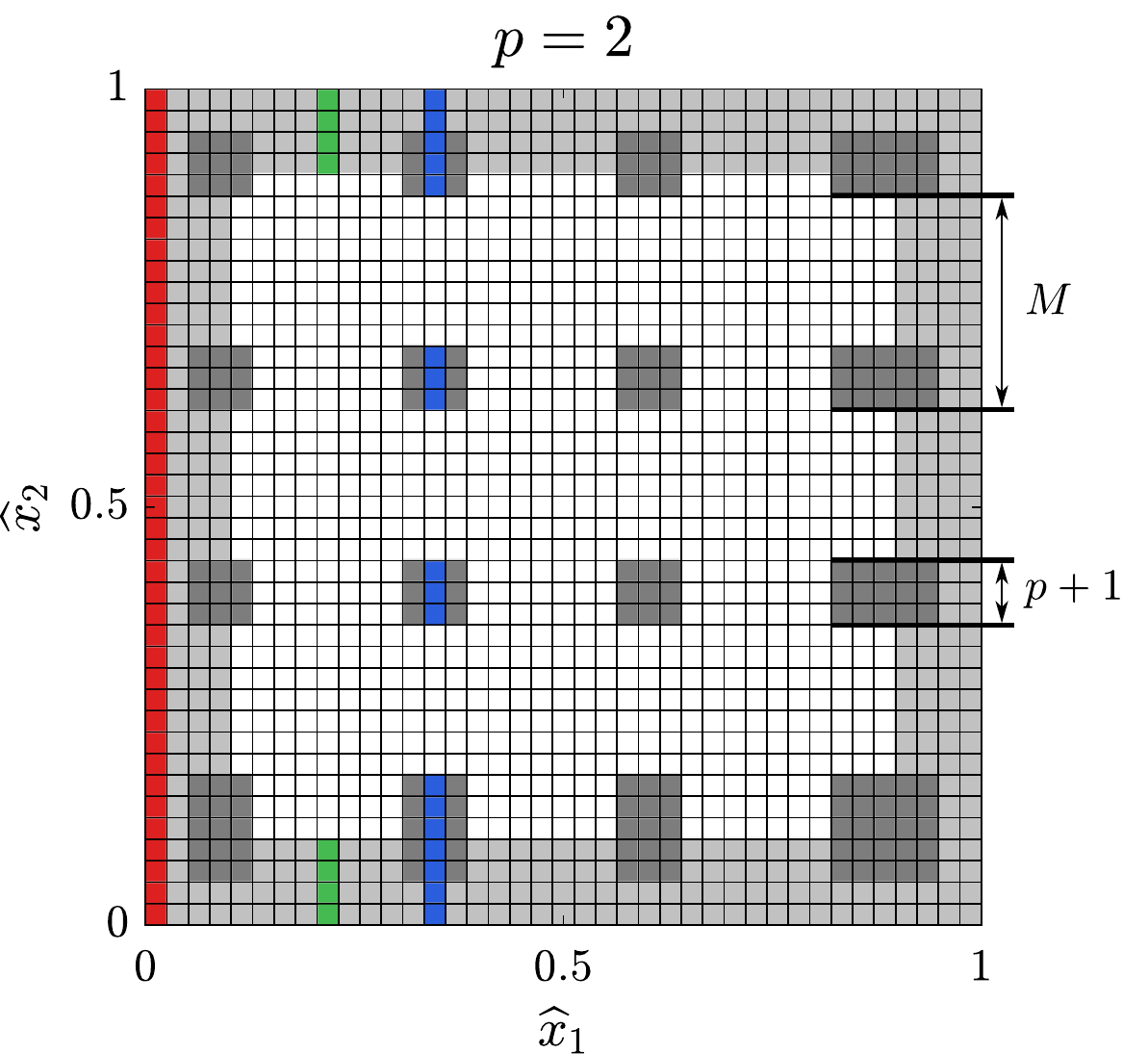}
\qquad\qquad
\includegraphics[height=5.5cm]{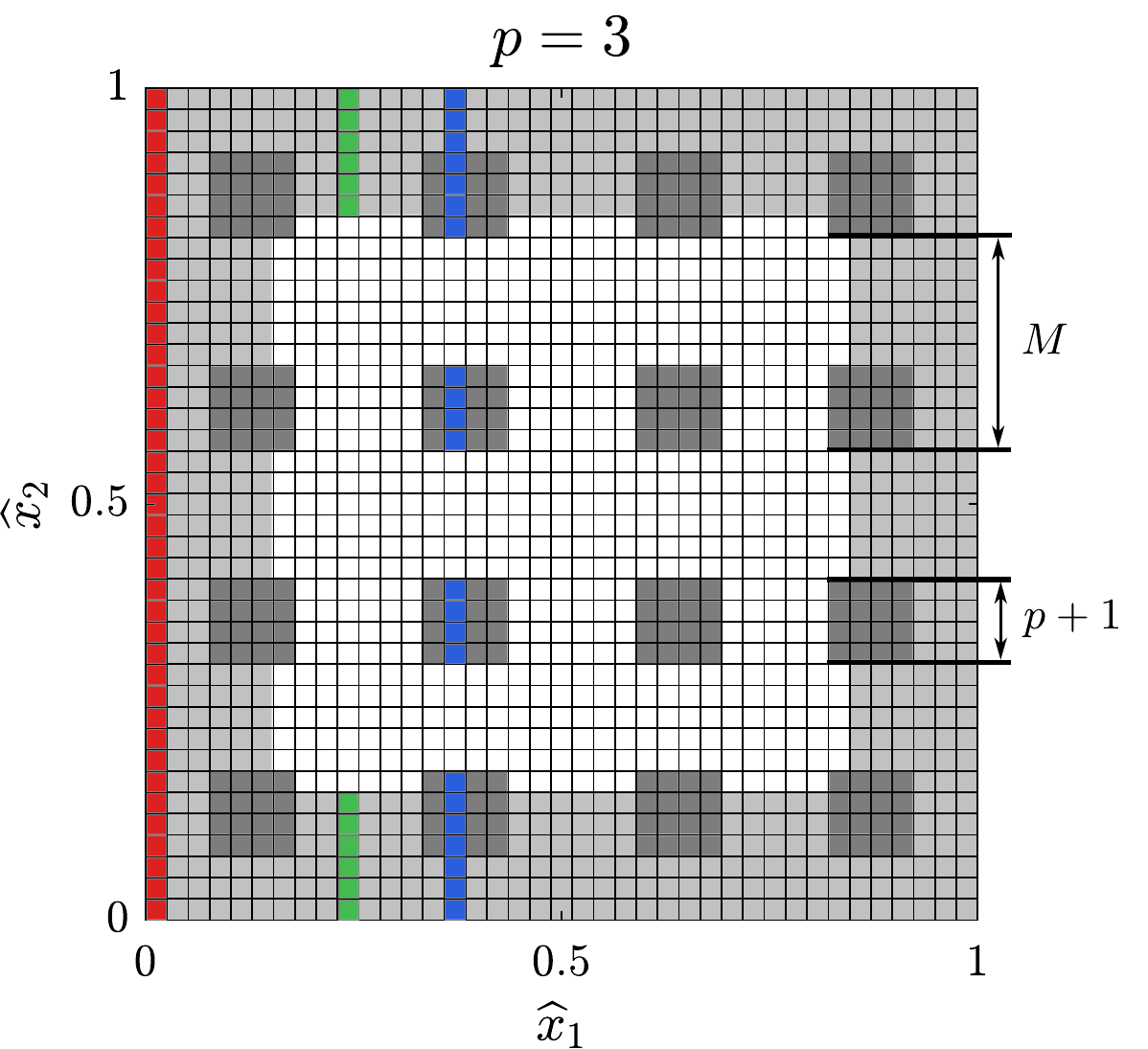}
\caption{\label{fig:ActiveElements}The active elements (shown in gray) involved in the surrogate assembly for $M=10$ with forty knots in each Cartesian direction. The light gray elements correspond to the active boundary elements and the dark gray elements correspond to the inner active elements required for the sampling of the stencil functions. The red, green, and blue elements correspond to the three different active element masks.}
\end{figure}

Let us denote the elements where quadrature in necessary as \emph{active elements}, cf.~\Cref{fig:ActiveElements}.
We added the possibility to perform quadrature on only a subset of the elements in a given column by providing a list of indices called a \emph{mask} for the active elements.
We implemented this feature by extending the core functions mentioned above by an additional optional argument named \texttt{element\_mask} which, as the name suggests, consists of a cell array with vectors of indices of active elements.
Particularly, the \texttt{element\_mask} is a cell array with a mask for the second dimension in the first component and, if required, contains an additional mask for the third dimension in the second component.

Some typical patterns of active elements are depicted in \Cref{fig:ActiveElements}.
Here, we have illustrated two different element masks corresponding to the cases $p=2,3$, $M=10$, and $N_\mathrm{el} = 39$.
Let us focus on the case $p=2$.
Here, the first and last four columns correspond to near-boundary elements.
That is, each whole column of elements is made up of active elements; cf.~the red column in \Cref{fig:ActiveElements}.
In this scenario, we should employ unmodified function calls to \texttt{msh\_evaluate\_col} and \texttt{sp\_evaluate\_col}.

Another scenario is when only the top and bottom near-boundary elements in a column are active; cf.~the green elements in \Cref{fig:ActiveElements}.
Here, the element mask is given by the set $\mathcal{B} = \{1,\ldots,2p\} \cup \{N_\mathrm{el}-(2p-1),\ldots,N_\mathrm{el}\}$, which yields the element mask $\{1,2,3,4,36,37,38,39\}$ for $p=2$.

In the third and final scenario, both elements in the interior and near the top and bottom boundary are active; cf.~the blue elements in \Cref{fig:ActiveElements}.
Each interior active element patch consists of $p+1$ elements and we start sampling at the $p+1$ element with an increment of $M$.
Since we want to avoid extrapolation in the subsequent spline interpolation, we also include the active elements near the interior boundary.
With $K = N_\mathrm{el} - 3p - 1$, the interior indices are thus given by $\mathcal{I} = \bigcup_{0 \leq k \leq \lfloor\frac{K}{M}\rfloor} \{kM + p + 1,\ldots, kM + 2p + 1 \} \cup \{K + p + 1, \ldots, K + 2p + 1\}$.
The final element mask is obtained by also including the near-boundary elements, i.e., $\mathcal{I} \cup \mathcal{B}$.
In our case $p=2$, the element mask is described by the set $\{1,2,3,4,5,13,14,15,23,24,25,33,34,35,36,37,38,39\}$.

\subsection*{Code extensions}
\label{sec:extensions}
Below is the signature of \texttt{op\_gradu\_gradv\_surrogate\_2d}. This function has the following input arguments: a \texttt{space} of type \texttt{sp\_scalar}, a \texttt{mesh} of type \texttt{msh\_cartesian}, a function handle \texttt{coeff} of the coefficient, the skip paramater \texttt{M} and the surrogate interpolation degree \texttt{q}.
The final surrogate matrix \texttt{K\_surr} (in sparse format) is the sole output.

\matlabfile[firstline=36,lastline=36]{src/op_gradu_gradv_surrogate_2d.m}
At the beginning of the function, some self-explanatory sanity checks are performed which verify that the correct input parameters have been passed.
\matlabfile[firstline=38,lastline=52]{src/op_gradu_gradv_surrogate_2d.m}
In the next few lines, some helper variables holding the number of degrees of freedom (dofs) in the univariate B-spline basis are defined.
The total number may be expressed as $N_\mathrm{el} + p$.
Removing $2p$ dofs from both the left and right boundary yields the number of interior dofs $N_\mathrm{el} - 3p$ present in one Cartesian direction in the stencil function domain.
\matlabfile[firstline=54,lastline=56]{src/op_gradu_gradv_surrogate_2d.m}
The following lines compute the indices of the rows in the global stiffness matrix corresponding to the stencil function domain.
This is done by collecting all available indices and removing the $4p$ indices nearest the boundary in each dimension.
In the 3D case, the array \texttt{row\_indices} has 3 dimensions.
\matlabfile[firstline=59,lastline=65]{src/op_gradu_gradv_surrogate_2d.m}
The next step is computing the sample point coordinates in the interior of the stencil function domain $\tilde{\Omega}$ described in \cite{drzisga2019igasurrogate}.
For practical reasons, we map this domain to the unit domain $[0,1]^n$ and take every $M{}^\text{th}$ point in each direction.
To avoid extrapolating any values later, we reinsert the sample points on boundary of $[0,1]^n$ which may have been skipped.
\matlabfile[firstline=68,lastline=72]{src/op_gradu_gradv_surrogate_2d.m}
Prior to computing the element mask, we save the indices of the matrix rows corresponding to the sample points \texttt{row\_indices\_subset}.
These indices are required in order to determine the active elements with the GeoPDEs function \texttt{sp\_get\_cells}.
In order to make the coming call to \texttt{sp\_get\_cells} faster, we filter the \texttt{row\_indices\_subset} to only include the rows up to index $(2p+1)\,(N_\mathrm{el} + p)$.
The last line filtering the rows is optional and may be omitted.
\matlabfile[firstline=75,lastline=77]{src/op_gradu_gradv_surrogate_2d.m}
The following snippet sets up an element mask which skips the quadrature at all non-active elements.
Here, we employ the \texttt{sp\_get\_cells} function which returns the indices of all elements within the support of the basis functions corresponding to the rows filtered in the previous snippet.
Some additional filtering enforces that only the indices up to $N_\mathrm{el}$ are included.
Since quadrature needs to be performed on all of the remaining near-boundary elements, we ensure that the elements with first or second indices in $\mathcal{B}$ are also active.
\matlabfile[firstline=80,lastline=84]{src/op_gradu_gradv_surrogate_2d.m}
Additionally, a second mask with only the near-boundary element indices is required.
Clearly, this mask only includes the indices in $\mathcal{B}$.
\matlabfile[firstline=87,lastline=88]{src/op_gradu_gradv_surrogate_2d.m}
For performance reasons, we pre-allocate memory for the sparse matrix with an estimated number of non-zero entries per row.
\matlabfile[firstline=91,lastline=91]{src/op_gradu_gradv_surrogate_2d.m}
Below is the first main loop which only performs quadrature on the active elements; i.e., those required for the subsequent stencil interpolation.
It is very similar to the original column-wise loop in \texttt{op\_gradu\_gradv\_tp}, but it employs the modified \texttt{msh\_evaluate\_col} and \texttt{sp\_evaluate\_col} (see previous subsection) with the previously determined masks.
\matlabfile[firstline=94,lastline=112]{src/op_gradu_gradv_surrogate_2d.m}
We define three vectors, which will hold the columns, rows, and values of the surrogate matrix.
This format allows for faster construction of a sparse matrix in a compressed sparse column format.
\matlabfile[firstline=114,lastline=116]{src/op_gradu_gradv_surrogate_2d.m}
The following snippet contains the main loop of the surrogate assembly algorithm.
In practice, there are $|\scD| = (2p+1)^n$ surrogate stencil functions $\tilde{\Phi}_\bdelta(\cdot)$, so the statements in the inner-most loop are reached $(2p+1)^n$ times.
First, the position (i.e., \texttt{shift}) of the stencil function relative to the diagonal entries in the matrix is computed.
This ``shift'' is in one-to-one correspondence with the translations $\bdelta\in\scD$ described in \cite{drzisga2019igasurrogate}.
Due to symmetry, we only need to compute the surrogates for the upper-diagonal matrix, thus we skip all cases where the shift is smaller or equal to zero by simply continuing to the next iteration.
If the shift is larger than zero, we extract all the sample points from the associated rows of the partly assembled stiffness matrix.
Having all the sample points $\tilde{\bmx}^\mathrm{s}_i$ at hand, we can easily perform the interpolation of the remaining matrix entries $\tilde{\Phi}_\bdelta(\tilde{\bmx}_i)$.
In order to remove any dependence on external software, the built-in Matlab function \texttt{interp2} is used here.
Note that this function only supports the spline interpolation orders $q=1$ and $q=3$.
In \cite{drzisga2019igasurrogate}, we used the \mbox{\texttt{RectBivariateSpline}} function provided by the SciPy Python package \cite{scipy}, which supports any spline interpolation up to order~$5$.
In the last three lines, the rows, columns, and values of the surrogate matrix are added to the vectors required for the sparse matrix creation at the end of the routine.
Note that the values of the upper-diagonal are added to the lower-diagonal in order to enforce symmetry.
\matlabfile[firstline=119,lastline=142]{src/op_gradu_gradv_surrogate_2d.m}
For performance reasons, we pre-allocate the diagonal of the surrogate stiffness matrix with dummy values which are overwritten later.
\matlabfile[firstline=145,lastline=147]{src/op_gradu_gradv_surrogate_2d.m}
In the penultimate step, we combine the matrix components obtained through interpolation with those obtained through standard quadrature.
This is done by creating a matrix \texttt{K\_interp} with only the interpolated components.
We then zero all of the entries in \texttt{K\_surr} which have non-zero values in \texttt{K\_interp} and finally sum the two matrices.
This may seem complicated at first, but it yields good performance because the sum of both sparse matrices may be computed efficiently.
\matlabfile[firstline=150,lastline=155]{src/op_gradu_gradv_surrogate_2d.m}
In the final step, the zero row-sum property is enforced by setting the diagonal value to the negative sum of the off-diagonal values in each row.
\matlabfile[firstline=158,lastline=158]{src/op_gradu_gradv_surrogate_2d.m}
\begin{rmk}
The \texttt{op\_gradu\_gradv\_surrogate\_3d} function is in many ways similar to its 2D counterpart.
For instance, clearly some of the 2D arrays need to be changed to 3D arrays, the 3D interpolation function \texttt{interp3} needs to be used, and the number of stencil functions increases.
However, performing quadrature only on the active elements is more difficult, since in the column-wise iteration GeoPDEs uses, each column is now made up of a matrix of elements (instead of a vector).
Therefore, a much more complicated set of element masks need to be used; see lines 98 to 115 in \fnurl{https://github.com/drzisga/geopdes/blob/v3.1-surrogate/geopdes/inst/space/\%40sp\_scalar/op\_gradu\_gradv\_surrogate\_3d.m}{\texttt{op\_gradu\_gradv\_surrogate\_3d.m}}.
\end{rmk}

\section*{Verification}
\label{sec:examples}
\begin{figure}	\includegraphics[width=0.5\textwidth]{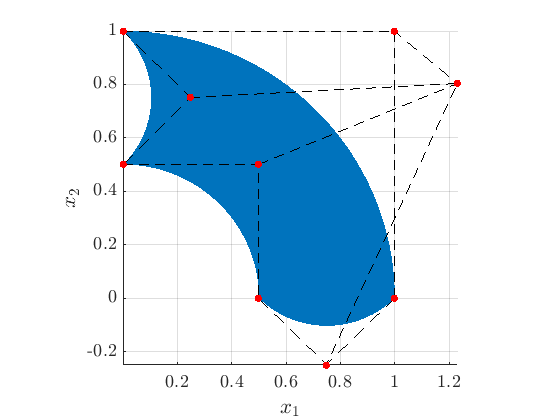}
	\includegraphics[width=0.5\textwidth]{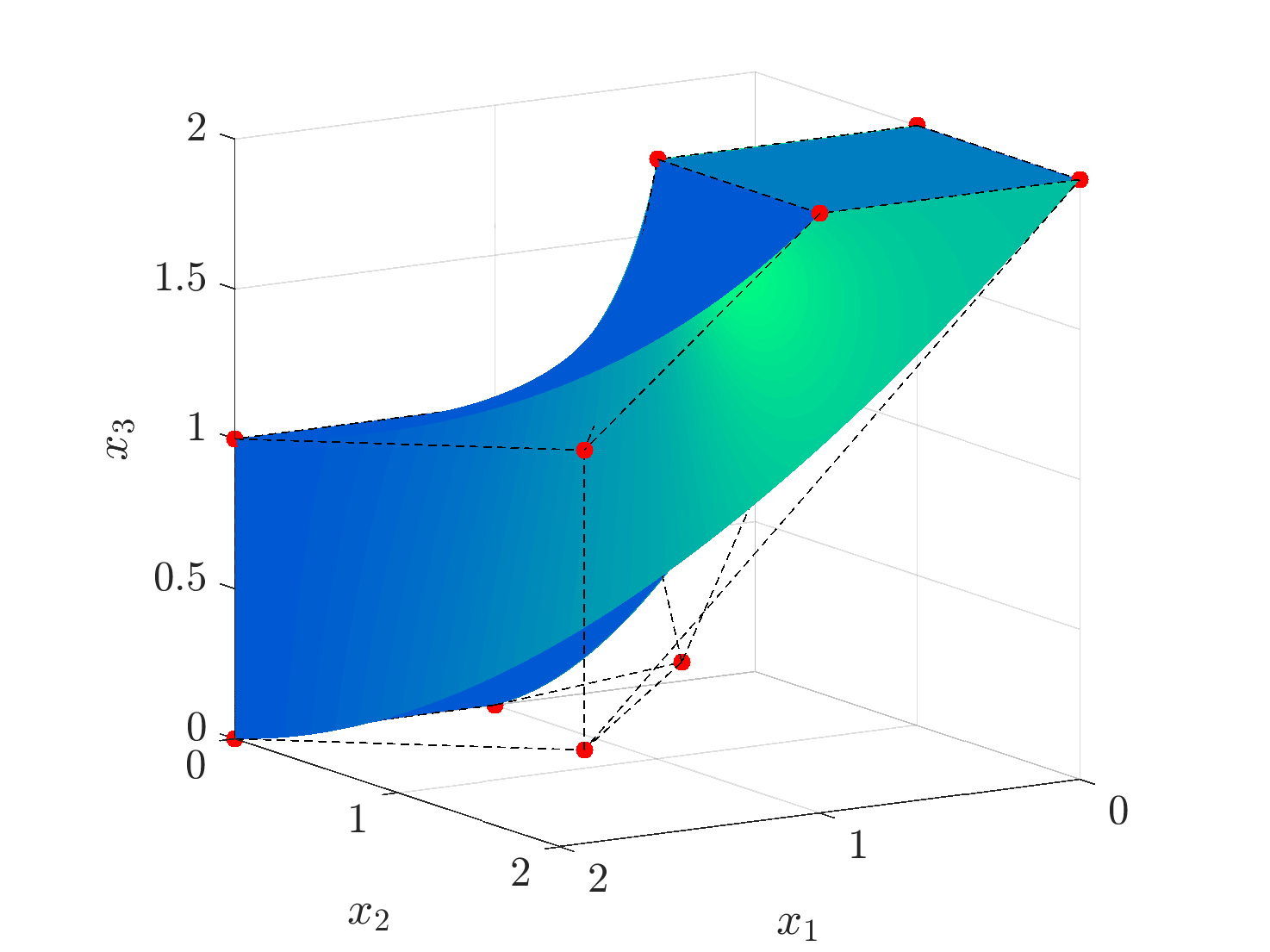}
	\caption{\label{fig:domains}Left: Domain $\Omega$ considered in the 2D example. Right: Domain $\Omega$ considered in the 3D example.}
\end{figure}

To facilitate easy verification, we added the demo script \fnurl{https://github.com/drzisga/geopdes/blob/v3.1-surrogate/geopdes/inst/examples/geopdes\_surrogate\_examples.m}{\texttt{geopdes\_surrogate\_examples.m}}, which provides the ability to run the problems considered in \cite[Section 7.3]{drzisga2019igasurrogate}.
The code for the 2D and 3D examples may be found in \fnurl{https://github.com/drzisga/geopdes/blob/v3.1-surrogate/geopdes/inst/examples/surrogate/ex\_surrogate\_poisson\_2d.m}{\texttt{ex\_surrogate\_poisson\_2d.m}} and \fnurl{https://github.com/drzisga/geopdes/blob/v3.1-surrogate/geopdes/inst/examples/surrogate/ex\_surrogate\_poisson\_3d.m}{\texttt{ex\_surrogate\_poisson\_3d.m}}, respectively.
Again, recall Poisson's problem
\begin{equation*}
\begin{alignedat}{2}
-\Delta u &= f &&\quad \text{in } \Omega,\\
u &= g &&\quad \text{on } \partial\Omega,
\end{alignedat}
\end{equation*}
on a domain $\Omega \subset \mathbb{R}^n$.
The 2D and 3D domains we considered are depicted in \Cref{fig:domains} and their respective NURBS/B-spline descriptions are in \fnurl{https://github.com/drzisga/geopdes/blob/v3.1-surrogate/geopdes/inst/examples/surrogate/data\_files/GeomQuarterAnnulusWithBumps.m}{\texttt{GeomQuarterAnnulusWithBumps.m}} and \fnurl{https://github.com/drzisga/geopdes/blob/v3.1-surrogate/geopdes/inst/examples/surrogate/data\_files/GeomBentTwistedBox.m}{\texttt{GeomBentTwistedBox.m}}.
In order to test the quality of the discrete solutions, we employ the following manufactured solution and load in 2D
\begin{align*}
u(x,y) &= \sin(20\,\pi\,x)\,\sin(20\,\pi\,y),\\
f(x,y) &= 800\,\pi^2\sin(20\,\pi\,x)\,\sin(20\,\pi\,y),
\end{align*}
and the following in 3D
\begin{align*}
u(x,y,z) &= \sin(20\,\pi\,x)\,\sin(20\,\pi\,y)\,\sin(20\,\pi\,z),\\
f(x,y,z) &= 1200\,\pi^2 \sin(20\,\pi\,x)\,\sin(20\,\pi\,y)\,\sin(20\,\pi\,z).
\end{align*}
The Dirichlet datum $g$ is chosen as the restriction of the manufactured solution to the boundary $\partial\Omega$.

Each example script assembles the standard IGA matrix $\sfA$ first and solves the corresponding system.
Afterwards, the surrogate matrix $\tilde{\sfA}$ is assembled and the surrogate system is solved.
Finally, the relative $L^2$ and $H^1$ errors for each case are computed and written to the console.
In addition, both the maximum absolute value of the difference in the two stiffness matrices (i.e., $\|\sfA-\tilde{\sfA}\|_{\max}$) and the achieved speed-up are displayed.

In the following, we present the output of running the \texttt{ex\_surrogate\_poisson\_2d} script in Matlab~2018b.
The script ran on a workstation equipped with an Intel\textsuperscript{\textregistered} Xeon\textsuperscript{\textregistered} E5-1630 v3 processor with a nominal frequency of 3.7 GHz.
\begin{minted}[fontsize=\footnotesize,frame=single,framerule=0.4pt]{text}
Initializing problem...
Assembling standard IGA matrix...
Standard assembly time: 5.022883 s
Solving standard IGA problem...
Standard solve time: 0.270706 s
Assembling surrogate IGA matrix...
Surrogate assembly time: 1.577257 s
Solving surrogate IGA problem...
Surrogate solve time: 0.273799 s
Computing errors...
Relative error in standard IGA
L2-norm: 1.335554e-03 
H1-norm: 1.407778e-02 

\|A-\tilde{A}\|_{\max} = 9.877796e-04

Relative error in surrogate IGA  
L2-norm: 1.335619e-03 
H1-norm: 1.407779e-02 

Assembly speed-up: 218.46\end{minted}

\begin{rmk}
The default parameters in the demo script are $p=2$, $q=3$, and $M=10$. The number of knots in each dimension is $160$ in 2D and $40$ in 3D.
These parameters may be easily modified by resetting the appropriate variables at the beginning of each script.
\end{rmk}

\section*{Acknowledgments}
This project has received funding from the European Union's Horizon 2020 research and innovation programme under grant agreement No 800898.
This work was supported by the German Research Foundation (DFG) and the Technical University of Munich (TUM) within the Priority Programme 1648 ``Software for Exascale Computing" (SPPEXA), by grant WO671/11-1, and in the framework of the Open Access Publishing Program.
\bibliographystyle{elsarticle-num-names}
\small
\setlength{\bibsep}{0.4pt}
\bibliography{../main_abbrv}
\end{document}